# The intermodal networks: a survey on intermodalism.

A. Merrina[1], A. Sparavigna[2] and R.A. Wolf[3]

[1] *Progetto Lagrange, Fondazione ISI, V.le Settimio Severo 65, 10133 Torino, Italy*
[2] *Dipartimento di Fisica, Politecnico di Torino, C.so Duca Abruzzi 24, 10129 Torino, Italy*
[3] *Business Development, Enercon Industries, Menomonee Falls, WI, USA*
(September 11, 2006)

A new and rather independent research field is now emerging on intermodalism, because intermodalism, that is the intermodal transportation, must study the behavior of systems different from single-mode transport systems. In fact, the importance of this new research field, its potential interdisciplinary of modeling the problem with the science of complex networks, are not yet completely appreciated by the scientific community. This is due to several reasons, among others the lack of a common background and a rather cumbersome approach of operation research to the modeling of transportation.
In this paper we review the concept of intermodalism and the approach to intermodalism in the operation research field. Then we discuss the intermodal networks in the framework of real world networks to understand the behavior of intermodal systems as a whole.

## 1. INTRODUCTION

Intermodal transportation, or intermodalism for short, is the natural answer to the requirements for moving people or cargoes. It can be viewed as a consequence of the existing social, marketing and distribution connections and, at the same time, as a source of new and more wider connections [1-3].
The term "intermodalism" gained currency when the marketing advantages of a transport network based on different size carriers with global alliances and strategies turned the transportation industry into an intermodal system - in particular when trailer-sized containers began to be widely used. Until the introduction of modern containerization, the process of transferring freights between different modes of transportation remained substantially unchanged. Containerization, that is the use of common containers with common shipping documents and tracking, uses a seamless transport network moving freight from origin to destination in a continuous pattern.

If the level of disjunction between modes is high, freight transportation can be described and modeled in terms of the separate modes, rather than by the common activities at the interface between modes. In fact, each mode of transportation has its different network model. But networks cannot be considered as independent when they are highly integrated and then new intermodal network models must be developed.

As pointed out by A. Donovan [4], the term "intermodalism" was at the beginning used to describe a transportation system composed of modes whose differences are more important than their common concerns. But today it is used not to emphasize the differences between the modes but rather a highly integrated system with different manifolds. The evolution of the contemporary industry of transportation into a comprehensive system is definitively moving beyond modalism to intermodalism.

Let us start with a survey of operation researches on intermodalism. Then we will discuss the most recent idea on modeling which networks the routing on freights and passengers.

## 2. AN EMERGING RESEARCH FIELD.

Studies on transportation systems are still under the domain of operations research. In this research area, the rather independent research field on intermodalism is emerging because the intermodal transportation system is intrinsically different from single-mode transport systems [5]. The importance of this new research field and its interdisciplinary, such as the possibility of modeling it with the science of complex systems and networks, is, in fact, not yet completely appreciated.

To be truly effective, research on intermodalism needs a common ground terminology that is still lacking. Each transport mode has its own terminology persisting in the operation of the current transportation system with limited coordination in the development of intermodalism [5,6]. The use of a different terminology hinders a faster development of integrated models.

Four modes of transportation carry freight and passengers: water, air, rail and road. Water transport moves bulk cargo and passengers at limited speeds to limited destinations. Air transport rapidly moves cargo and passengers in limited quantities and numbers to limited destinations. Rail transport moves large quantities of cargo over land routes to limited destinations. Road transport moves cargo and passengers to virtually any destination in limited quantities. Each mode is developed independently. And if the connections among the modes is limited, each transportation mode has its own network model. From the above description of mode characteristics, it is evident that air and water transport have networks with a limited low number of nodes but high number of connections among modes - different from rail and road transports (Fig.1). The penetration of these two kinds of networks in the intermodal system gives origin to new studies [7]. An intermodal transportation system is a collection of passengers and cargo moving via multiple modes of transportation. The routes along which they are moved form the connections of the network. The network nodes are the terminals at which they are stored, transferred, etc.

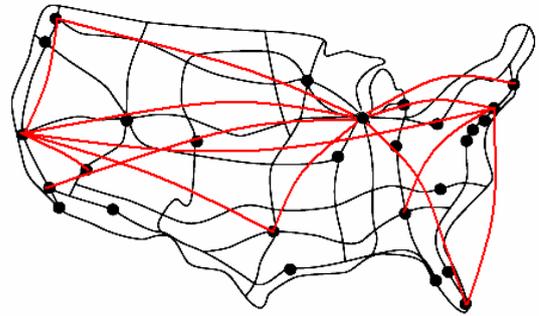

FIG.1: Rail- (in black) and air- (in red) networks display a different structure of connections among nodes.

To help achieve the potential benefits of an intermodal transportation system, modelers attack systems analysis and design problems using computer simulation, queuing analysis, mathematical programming, probabilistic analysis, graphical analysis and other approaches [8]. Usually, it is the operations research, a scientific discipline dealing with the problems of methods for decision making, that is developing models for transportation. In [8,9], literature reviews of mathematical modeling of intermodal transportation are given. Ship terminal models dominate the intermodal freight literature [10-13]. In [10], a ship-to-rail intermodal freight terminal is simulated using the physical elements of terminal and the terminal operations processes of loading and unloading cargo. Frequency and time of arrival and departures, economic costs and required operations are functions to optimize.
Models of intermodal passenger facilities are displayed in Refs. [14-17]. In [14], for instance, an intermodal train terminal is simulated with personal and public vehicles, and regional and local buses. Comparative evaluation of conventional and advanced rail-road terminal equipment is shown in Ref.18. Statistical or deterministic inputs have been used in [19] to simulate the flow among intermodal terminal units in a rail-road system where intermodal terminals are connected by rail corridors. From a survey of the literature, it is evident that the operation research approach is not discussing the statistic nature of the network involved in transportation.

## 3. PERFORMANCES, ISSUES AND COMPLEXITIES OF MODELS

From an operations research approach to intermodal systems, the intermodal system is viewed as a collection of items that act together toward the accomplishment of some end. A model of the system is developed as a simplified representation of a system. Besides iconic and analog models, symbolic model are used to analyze the behavior of the system. Iconic models are scaled models of the actual (real) systems and analog models are physical models used to describe the original system (for instance, the use of electrical circuits to model thermal conductivity). Symbolic models are often referred to as mathematical models, because the system behavior is described using only equations and logical relationships [20,21].
In these models, variables and parameters must be carefully chosen. Variables, or decision variables, are those quantities over which the system manager has a control, often limited by constraints. Parameters are values over which the decision-maker has no control. Performance measures are quantities that capture the level to which the system is operating. Examples of performance measures are equipment utilization, operating costs, and so on. An objective function identifies an important performance measure and the optimization goal (maximize or minimize) for the measure. For example, an objective function may minimize operating costs or maximize the profit. In a mathematical model, decision variables, parameters, constraints, performance measures, and objective functions are all captured using equations and/or logical relationships. The transportation terms are needed to complete the terminology base for modeling intermodal transportation systems.
There are many issues that must be considered when addressing intermodal transportation routing. First of all, the fixed and variable transportation costs are different in each transportation mode. In the intermodalism, the model must consider not only the cost for each different mode, but also the transfer cost from mode to mode and the transfer time. Transfer costs depend upon the transfer point at which they occur [22]. In the model then, costs and times are represented by means of functions which must be minimized of maximized.
When modeling intermodal transportation systems, various factors can affect the representation of the reality of the system. One factor to be investigated by a modeler is the consideration of single or multiple routes between shipment origin and destination. With each additional path considered, the modeler must face a large increase in the complexity of model behavior. Assuming that there are a number of modes available at each city in the transportation network, transport of passengers or freights from the starting point to the destination will have several possible paths on a network.
The main issue of an intermodal transportation lies in minimizing the overall transportation cost. This cost can be estimated adding the equipment and crew costs, overhead costs, and general and administration costs. Sometimes, multiple objectives must be included in modeling. That is, for instance, not only minimizing total cost, but also minimizing total time and/or maximizing the service level and social benefits. A possible approach is to weight the different objectives based on their importance [23-25]. One of the objectives of operation research is developing methodologies to select the best combination of transportation modes to move a shipment from an origin to destination, including transfer costs between modes. The goal is to obtain a decision support system for real-time routing of shipments through an intermodal network.

A report about the studies on transportation problems with operative programs is rather wide and dispersive. We prefer avoiding deeper descriptions of these studies and devote our attention to networks, starting with a simple problem related to transportation; that is, the problem of the shortest path.

## 4. THE SHORTEST PATH PROBLEMS.

Fixing our attention on the network, the primary problem is to find connections between the nodes of the network to determine possible paths from origin to destination and to define costs functions for each path. In the network, cities are represented by nodes, and routes are represented by links between the nodes. The distances of these links represent transportation cost or time as opposed to the traditional linear distance in miles. Single or multiple routes between origin and destination are possible. Each additional node or path increases the complexity of the model.

One of the main problems in the modeling and analysis of transportation and communication networks is to find the shortest path for the lowest cost of transportation. It could desirable, however, to determine not only the shortest path through a network, but to identify several sets of paths on the network where the choice is dependent upon multiple objectives or special circumstances [26-28].

In the modeling of transportation involving multiple modes between each node of the network, two manners of approach are possible. One way is to represent a node in the network once for each mode of transportation that can enter that city. For example, if a city has three modes of transportation available to enter the city, the city is represented by three nodes. Then, there is only one link between each pair of nodes in this modeling method. Each link corresponds to the transportation cost of the mode represented by the node to which it is linked. The transfer cost from one type of node to another is represented in this method with a link. It is a link leaving a node designated as one transportation type, and entering a node designated as another. These kind of links represent transfer costs.

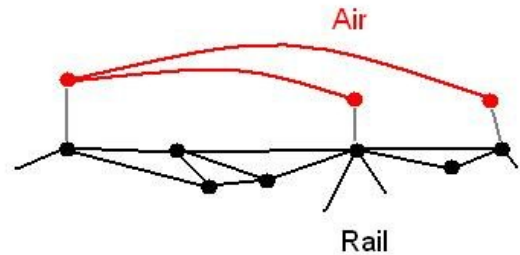

FIG.2: A model of intermodal transportation, where a city is represented with a node for each transport mode entering the city. In gray, the transfer links from mode to mode.

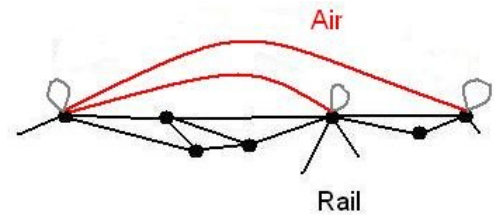

FIG.3: Multiple links represent the modes entering the city. The gray loops are the transfer links.

The multiple link method of transportation network modeling involves representing each city by one node and allowing more than one link between any two nodes. Each transportation mode is represented by a link between cities. One and only one node represents each city. Each link contains the transportation cost of the mode it represents. This type of model is smaller by definition than the multiple node method, but in this case the transfer cost from one type of node to another must be represented by loop on the node (a link that leaves a node and enters the same node).

A starting and an ending node representing the origin and destination are placed on the network. The analysis of paths from origin to destination allows for cost calculations. The shortest path methodology needs one specific node at which to start the shortest path calculations, and one specific node at which the calculated paths should end. This is what is called the problem of the Origin-Destination OD shortest path. The analysis of transportation networks, in which the computation of shortest paths is the most fundamental problem, have been the subject of extensive research for many years [29-31].

What could be interesting for further development of an intermodal network is an overall evaluation of the efficiency of the network itself. This can be obtained by adapting the real-world network formalisms to intermodalism.

## 5. SMALL WORLDS.

What is happening now in the transportation system is similar to what is observed in the development of the World Wide Web. Until recently, all researches devoted to internet development concentrated on improving the communication protocols. Now it is starting a deep investigation on what exactly the web is and what could be in the future the WWW under development. A. Barabási pointed out that the behavior of the web is appearing much more as an ecological system than a huge electronic system and that, to understand the WWW nature and evolution, it requires much more attention to the topology of the network behind it [32,33].

The pervasiveness of networks is evident: networks are the internet, the power grids, and the transportation systems, such as human societies and ecosystems. Finding models which describe and predict the global behavior of networks is then, quite appealing for scientists. Nevertheless, in spite of the importance of networks, the research for understanding their structures and properties is just the beginning.

In this section, we want to show the most important properties of real-world networks. Of course, the knowledge of such properties is of fundamental importance in the development of accurate models.

The modeling of networks is rather simple and natural. It can be represented with a graph with nodes and links connecting nodes. Social networks have been the first complex networks explored. One of the most famous experiments in social systems was performed by S. Milgram in 1960. This small world experiment showed that the average number of steps in an acquaintance chain length was only about six [34]. Analyses on other networks showed a similar property, that it is possible to reach a node from another one by going through a number of links, small if compared to the total number of existing nodes [35-40].

All the functions defining the performances of networks modeled with graphs, are built with numbers of nodes and links and with the lengths of the links. For instance, to measure the typical separation between two generic nodes in a graph G, the characteristic path length L coefficient was introduced [36]:

$$L(G) = \frac{1}{N \cdot (N-1)} \cdot \sum_{\substack{i,j \in G \\ i \neq j}} d_{ij} \qquad (1)$$

where N is the total number of nodes in G and $d_{ij}$ the shortest path length between nodes i and j (i.e. the minimum number of links covered to reach j from i). For a theory considering non-connected graphs as well, a global efficiency can be defined [41]. Efficiency between nodes is inversely proportional to the shortest path length. When the graph is not connected, efficiency is equal to 0.

A high clustering is another important property of real–world networks [36]. To estimate the clustering of a graph G, a clustering coefficient C is defined as follows. For each i-node, a subgraph $G_i$ of G is considered, where the nodes are just the first neighbors of the i-node. The i-node and all the links incident in it are removed. If the node i has $k_i$ neighbors, $G_i$ will have $k_i$ nodes and at most $k_i(k_i-1)/2$ links. $C_i$ is the fraction of links actually existing and C is the average of $C_i$, calculated over all nodes:

$$C(G) = \frac{1}{N} \cdot \sum_{i \in G} C_i \qquad (2)$$

where

$$C_i = \frac{\text{Number of links in } G_i}{k_i \cdot (k_i - 1)/2}. \qquad (3)$$

Several definitions of C are present in literature [42]. Global and local scales of the system are therefore analyzed by means of these two parameters. Let us note that a "high clustering" of a network means an high efficient system on a local scale.

Another important parameter is the degree of a node, defined as the number of links incident with the node. Consistently the distribution P(k) of degrees is defined as the probability of finding nodes with k links: P(k)=N(k)/N, where N(k) is the number of nodes with k links. Many very large networks have the distribution following a power law for large k:

$$P(k) \sim k^{-\gamma} \qquad (4)$$

where γ is an exponent that often varies between 2 and 3. Networks with such distribution are scale-free networks.

As we have seen in the previous sections, in real-world networks each link has its cost. Transportation systems are not interested in realizing high cost networks, i.e. networks with a large number of links. The same is in the nature networks minimizing the free energy. To estimate the cost of a graph G, with N nodes and K links, the following normalized number of links can be used [43]:

$$\text{Cost}(G) = \frac{2 \cdot K}{N \cdot (N-1)}, \qquad (5)$$

where cost varies in the range [0,1].

It is possible to define some simple special networks, but these are rather far from the model of real networks. One is the regular network, where each node has the same number of links. Because of its excessive regularity, this model is a bad representation of reality. Though this network can possess a high clustering, it lacks good global properties (L~N/2k>>1). In the Cayley tree (or Bethe lattice), each node has k links as in the regular lattice, but no cycles exist. Such a network can simply explain the phenomenon of the "six degrees of separation" law. The minimum number of steps necessary to reach a generic node from the root, that is the origin of the tree, is at most D=log(N)/log(k). If we assume in the world a population of about $N=10^9$ and $k=10^2$, we find that 4.5 average steps are enough to reach any person from the root.

6. RANDOM NETWORKS.

Regural or Bethe lattices are not able to describe complex systems. In 1959, two Hungarian mathematicians, Paul Erdôs and Alfréd Rényi, proposed to randomly build the networks used to describe communication and life systems [44,45]. Their recipe was simple: a random graph can be constructed from an initial condition with N nodes and no connections by adding K edges (in average k per node), randomly selecting the couples of nodes to be connected. This model and the related theorems proposed by Erdôs and Rényi revitalized the graph theory, leading to the emergence of a new research field focusing on random networks [46].

In a random network the nodes follow a Poisson distribution with a bell shape, with a low probability to find nodes that have significantly more or fewer links than the average node (see Fig.4). Random networks are also called "exponential networks" because the probability that a node is connected to *k* other nodes decreases exponentially for large *k*.

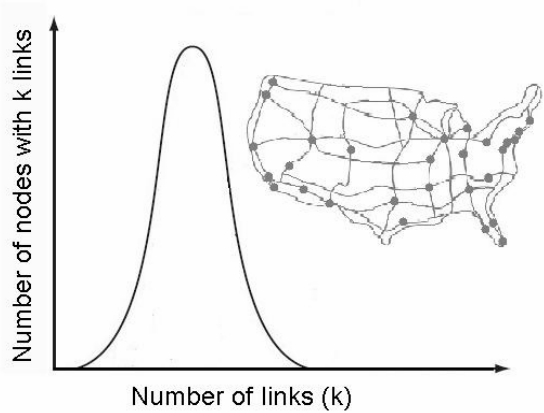

FIG.4: A random network has a bell shape distribution.

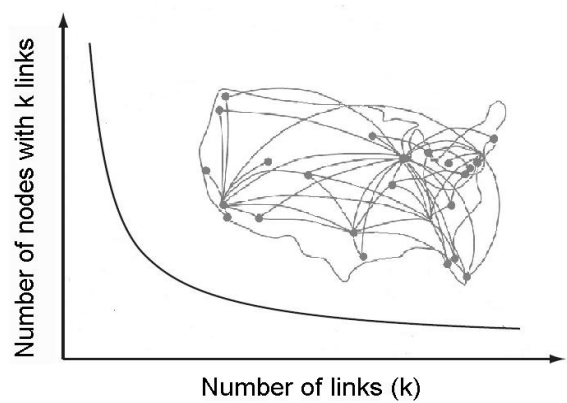

FIG.5: A free-scale network with a power law distribution

If low cost is assumed, the Erdös-Rényi model presents a small characteristic path length (L=log(N)/log(k)), but low clustering (C=k/N) with a Poissonian degree distribution.

A random network could model a highway network, in which nodes are the cities, and links are the major highways connecting them. Indeed, most cities are served by roughly the same number of highways. In contrast, if we consider the air traffic system, we see that most nodes have only few links (Fig.5, in the Appendix, an example from history). The network is held together by a low number of highly connected hubs. A large number of small airports are connected to each other via these major hubs. In this case, it is a power law degree distribution which describes a scale-free network predicting where most nodes have only a few links. This was discovered by A. Barabási and his colleagues at the University of Notre Dame in Indiana, when in 1999 they mapped the connections of the Web. The Web did not have an even distribution of connectivity, that is, a random connectivity, but rather a scale-free distribution. Barabási and his collaborators called the highly connected nodes "hubs" and coined the term "scale-free network" to describe the class of networks that exhibit a power-law degree distribution which they presumed to describe all real-world networks of interest.

It was shown later that most of the real-world networks can be classified into two large categories according to the decay of P(k) for large k [47].

Improvements on the Erdös-Rényi model were proposed by Watts and Strogatz [36] and by by Molloy and Reed [48]. The model of Watts and Strogatz is interpolating between regular and random graphs, with few long range connections leading to good global properties, without altering the high clustering, typical of regular networks. The model proposed by Molloy and Reed realizes a random graphs with a generic degree distribution with small path length and low clustering. Though the model is able to generate networks with a scale-free degree distribution, it does not displays the dynamical evolution of real-world networks.

To mimic the dynamics of real-world networks, a network needs to grow and attach. Most real-world networks are open systems, in which the number of nodes (or links) is not fixed. Moreover, the new nodes are not randomly connected to existing nodes with uniform probability, but they are attached with greater likelihood to hubs. According to Barabási, the implementation of these two mechanisms in a model is enough to generate of scale-free networks.

The dynamics of the Barabási and Albert (BA) model starts with a small number of nodes $m_0$ and at each time-step a new node with $m \leq m_0$ links is created. With a preferential attachment, the probability that the new node is connected with the i-node is proportional to i-node degree $k_i$. Numerical simulations and analytical solutions of the model in the mean field approximation predicts a degree asymptotic distribution for $t \rightarrow \infty$:

$$P(k) = 2m^2 k^{-\gamma} \qquad (6)$$

with an exponent $\gamma = 3$, independent of m and of the size of the network. The BA network has connectivity properties ensuring a good efficiency in communication between nodes. The drawback of the model is its high vulnerability to attacks, because it is enough to remove few "hubs" and the efficiency of the network is destroyed [49]. Moreover, the exponent $\gamma$ is fixed and cannot be tuned to simulate different real networks.

An improvement of BA model was developed by Klemm and Eguìluz [50,51], which gives low cost scale-free networks, both globally and locally efficient. The model is based on the fact that nodes can age and that their state can be switched from the active to the non-active one, meaning for instance that they loose importance in time. Concerning attack tolerance, the Klemm-Eguìluz model shows results that are similar to those obtained for the BA model. Good representations of real-world networks are possible with this model, however, other high clustered scale-free models have been recently developed [52,53].

## 6. INHOMOGENEOUS NETWORKS.

To study the behavior of a network during its growth and its stability under attacks is clearly important. This last point is the main problem of passenger transportation after the interactivity and connectivity of the global transportation system was altered by terrorism. The models of the network previously discusses are vulnerable when the most important nodes of the network are attached. In the case of intermodalism, two or more modal networks are strongly interconnected, with hubs usually located in the same geographic area. A failure in one of these area provokes a decrease in the efficiency of both modes.

If we consider the road–air intermodal transport, the network model must be able to describe the random network properties of the road mode and the free-scale properties of the air mode. To the authors' knowledge, such a complex network model has not jet been developed, but models that seem more promising to be modified for describing the intermodalism are those belonging to the group of inhomogeneous networks.

The model of an inhomogeneous network is usually consisting of two types of nodes denoted by A and B. Type A nodes can be linked and link to other nodes, while type B nodes can only link to others but not allowed to be linked. That means links between two type B nodes are prohibited. At each step in generating the network, a new node (A or B) is created with probability p or 1-p respectively. The connectivity $k_i$ of the i-node is defined as the total number of links pointing to it. The presence of nodes with a very large number of connections is the key characteristic of these networks.

The probability P(k) that a node has k incoming links is following a power law $P(k) \sim k^{-\gamma}$, with $\gamma$ comprised between 2 and $\infty$. The network's parameter $\gamma$ and the centralization index of inhomogeneous systems are much closer to the networks in reality than the BA model. The centralization parameter, used to measure the network centrality quantitatively [42], for the BA network is about 0.0021, while the value of the internet WWW is about 0.02, about 10 times larger than that of BA model. In an inhomogeneous network, when probability p increases from 0.1 to 1, the centralization index changes from about 0.1 to the value of the BA model.

This is way the study of stability of inhomogeneous networks may reveal some characteristics of the real-life networks. The strategy of an intentional attack could be to destroy the nodes which have the most links. These nodes turn out to be removed with their links from the network. The measure of the possible damage provoked in the network can then be obtained by evaluating the local clustering (Eq.3) of the network and the average size of the isolated clusters [54,55]. Cohen et al. have reported a method to calculate the critical fraction of nodes needed to be removed in the sabotage strategy to disrupt the network [56].

Compared with a BA model, the inhomogeneous networks are much more fragile under sabotages, because the size of the clusters decreases much faster than in the BA network [55]. But, as shown by simulations of WWW and Internet networks under intentional attack, these networks broke down more quickly than BA model. When compared with the BA model then, the inhomogeneous model describes the realistic networks better.

## 7. CONCLUSIONS

We have deeply discussed the most recent point of view of the complex system science about networks and models. We did so to stimulate a more cross-correlation between this research area and the operation research field. Focusing exclusively on network modeling without regard for larger strategic ramifications will most likely result in sub-optimal results, and this is a quite general problem solvable only with a strong interdisciplinary approach. Logistics professionals, the end-users of the research results, mostly rely on modeling tools as decision support, not as determinants of the "perfect" distribution solution. Only by blending such theoretical tools into a deep understanding of the company's present and future business strategy and practical implementation requirements, can a network optimization generate real logistics cost savings.

## APPENDIX

In the history, an example of complex transportation network was the system of Roman roads. The network has Rome at a center from which roads are fanning out from it. With the growing of the empire, other cities become hubs of the system.

In fact, this network has a structure more similar to a modern air transport system, with few hubs and links of rather different lengths. Essential for the growth of the empire, by enabling it to move armies, this network was fundamental in maintaining both the stability and expansion of the empire.

Was the roman road system a free-scale network? It is not simply to answer. A possible starting point for approaching such a problem is to understand how the Romans displayed their road system. The Romans, as the ancient travelers in general, have no maps. However, to have an idea about runs and distances, they used "itineraria". In origin, itineraria were simply lists of cities along the roads. To sort out the lists, the Romans drew parallel lines showing the branches of the roads. On the lines they placed symbols for cities, way stations, water courses, and so on. Because the itinerarium does not represent landform, it is not a map, it is a graph. A beautiful example of an ancient graph is the Tabula Peutingeriana, one of the master itineraries showing the road network which covers Europe, parts of Asia and North-Africa.